\documentclass[11pt,twoside]{article}

\usepackage{asp2006}
\usepackage{epsf}
\usepackage{graphicx}
\usepackage{lscape}

\begin{document}
\title{The obscured quasar population from optical, mid-infrared, and X-ray surveys}
\author{C. Vignali}
\affil{Dipartimento di Astronomia, Universit\`a degli Studi di Bologna, Italy}
\affil{INAF -- Osservatorio Astronomico di Bologna, Italy}
\author{A. Comastri}
\affil{INAF -- Osservatorio Astronomico di Bologna, Italy}
\author{D.~M. Alexander}
\affil{Department of Physics, Durham University, UK}

\begin{abstract}
Over the last few years, optical, mid-infrared and X-ray surveys have brought to light a significant 
number of candidate obscured AGN and, among them, many Type~2 quasars, the long-sought after 
``big cousins'' of local Seyfert~2 galaxies. However, despite the large amount of multi-wavelength 
data currently available, a proper census and a panchromatic view of the obscured AGN/quasar 
population are still missing, mainly due to observational limitations. 
Here we provide a review of recent results on the identification of obscured AGN, 
focusing primarily on the population of 
Type~2 quasars selected in the optical band from the Sloan Digital Sky Survey. 
\end{abstract}

\section{Introduction: the X-ray and mid-infrared hunt for obscured quasars}
The quest for the identification of luminous and obscured Active Galactic Nuclei (AGN), 
the so-called Type~2 quasars predicted by unification schemes of AGN (e.g., Antonucci 1993) 
and required by many synthesis models of the X-ray background (XRB; e.g., Gilli, Comastri 
\& Hasinger 2007), has been the topic of numerous investigations over the past few years. 
Although moderate-depth and ultra-deep X-ray surveys (see Brandt \& Hasinger 2005 for a review) 
have proven effective to reveal a large number of Type~2 quasar candidates at high redshift 
(e.g., Alexander et al. 2001; Mainieri et al. 2002; Mignoli et al. 2004), 
the optical counterparts of these sources 
are typically faint and therefore represent challenging targets to obtain spectroscopic redshifts 
and reliable classifications based on ``standard'' optical emission-line ratio techniques. 
Furthermore, there is evidence that despite the fact that $\approx$~80\% of the XRB 
has been resolved into discrete sources by ultra-deep \hbox{X-ray} surveys in the 2--8 keV band 
(e.g., Bauer et al. 2004; Hickox \& Markevitch 2006), only $\approx$~60\% of the XRB 
has been resolved above $\approx$~6~keV (Worsley et al. 2004, 2005), 
indicating that while ultra-deep \hbox{X-ray} 
surveys provide efficient identification of AGN activity, they do not 
provide a complete census of the obscured AGN population. 
It is plausible that one of the ``missing'' XRB components is related to the population of 
Compton-thick AGN (i.e., sources with column densities $>1/\sigma_{\rm T}\approx10^{24}$~cm$^{-2}$, 
where $\sigma_{\rm T}$ is the Thomson cross section; see Comastri 2004 for a review on 
Compton-thick AGN) which are required by the AGN synthesis models of the XRB 
(Gilli et al. 2007) but, as being heavily obscured, are difficult to discover and identify 
in the X-rays by the current generation of X-ray telescopes; see, e.g., Tozzi et al. (2006) for the 
Compton-thick AGN candidate selection in the {\it Chandra} Deep Field-South. 

In the presence of obscuration, the nuclear emission is expected to be re-emitted at 
longer wavelengths and hence mid-infrared (MIR) observations can be crucial 
to reveal obscured AGN emission. 
Recently, numerous attempts have been made in this direction, fully exploiting the capabilities of 
the detectors on-board {\it Spitzer} (e.g., Alonso-Herrero et al. 2006; Polletta et al. 2006; 
Donley et al. 2007). 
Unfortunately, the AGN locus in the color-color diagrams obtained from {\it Spitzer} photometry 
(e.g., Lacy et al. 2004; Stern et al. 2005) is often contaminated by starburst galaxies, 
therefore further investigations and adjustments are required to efficiently distinguish the 
obscured and elusive AGN from the less intriguing unobscured population. 
%
%
The present limitations of {\it Spitzer} diagnostic diagrams to select obscured AGN may be 
overcome either by refining the selection criteria (e.g., Mart{\'{\i}}nez-Sansigre et al. 2005, 
2006) or targeting the optically faint or invisible source population with, e.g., MIR spectroscopy 
(e.g., Houck et al. 2005; Weedman et al. 2006a). 

The next obvious step where most of the observational efforts will be concentrated in the years 
to come is to compare the obscured AGN selection criteria adopted at different wavelengths 
and compute their efficiency in the detection and identification of the most heavily obscured 
quasars. 
Parallel to these kinds of studies, the analyses of the multi-wavelength (from MIR to X-rays) 
properties of obscured AGN and quasars will be crucial to investigate their emission 
accurately (e.g., Weedman et al. 2006b), refine the current torus models and templates (e.g., 
Silva, Maiolino \& Granato 2004), and derive some fundamental parameters 
such as the masses of the super-massive black holes (SMBHs) residing in these sources and 
their Eddington ratios (e.g., Pozzi et al. 2007). 
Finally, the co-evolution of galaxies and SMBHs will be investigated up to high redshifts, 
where a significant fraction of MIR--submillimeter-selected obscured AGN will probably be found 
(e.g., Alexander et al. 2005a; Mart{\'{\i}}nez-Sansigre et al. 2005).

\section{Are any optically selected Type~2 quasars out there?}
Although primarily designed, in the AGN research field, for the discovery of broad-line (Type~1) 
objects, the Sloan Digital Sky Survey (SDSS; York et al. 2000) has provided a sample 
of 291 high-ionization narrow emission-line AGN in the redshift range \hbox{0.3--0.83} 
(Zakamska et al. 2003; small filled circles in Fig.~\ref{lxoiiiz}), 
many of which are identified as candidate Type 2 quasars on the basis of their 
\hbox{[O\ {\sc iii}]}5007\AA\ luminosities. 
In the following, we summarize the main results obtained over the last three years 
for a sub-sample of these optically selected Type~2 quasars with {\it ROSAT}, {\it Chandra}, 
and XMM-{\it Newton} observations.
\begin{list}{$\surd$}{} 
\item 
From the analysis of primarily archival {\it ROSAT} observations, 
Vignali, Alexander \& Comastri (2004, hereafter V04) were able to place constraints 
on the X-ray emission of 17 SDSS Type~2 quasar candidates (open circles in Fig.~\ref{lxoiiiz}). 
Using the \hbox{[O\ {\sc iii}}] line luminosity to predict the intrinsic X-ray power of 
the AGN (following the correlation of Mulchaey et al. 1994), V04 found that at least 
47\% of the observed sample shows indications of X-ray absorption, including the four highest 
luminosity sources with predicted unobscured luminosities of 
\hbox{$\approx$~10$^{45}$~erg~s$^{-1}$}, 
hence well above the typically adopted threshold of 10$^{44}$~erg~s$^{-1}$ in the 2--10~keV band 
for Type~2 quasars. 

\begin{figure}
\centering
\includegraphics[angle=0,width=0.6\textwidth]{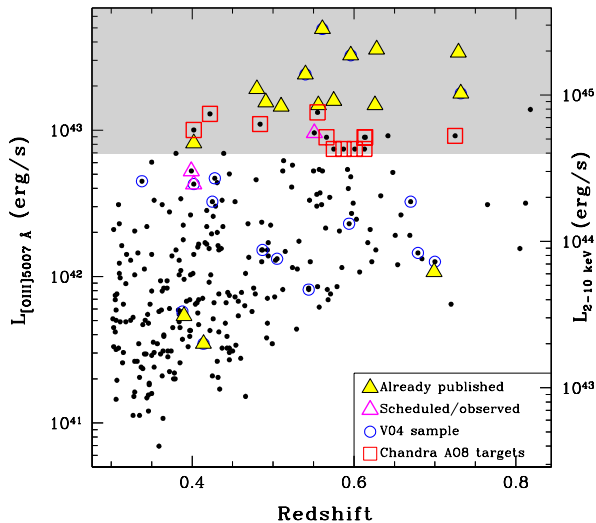}
\caption{
Logarithm of the measured $L_{[O\ III]}$ luminosities vs. redshifts for all of the sources in 
the original sample of Type~2 quasar candidates from Zakamska et al. (2003; small filled circles). 
At the right of the panel, the \hbox{2--10~keV} luminosities, 
estimated using the correlation between the \hbox{[O\ {\sc iii}}] and \hbox{2--10~keV} flux 
(Mulchaey et al. 1994), are shown. 
The key provides a description of the X-ray observations. 
The grey region defines the locus of the sources which still lie in the quasar regime 
(i.e., above 10$^{44}$~erg~s$^{-1}$) even taking into account the dispersion in the 
$L_{[O\ III]}$--$L_{\rm 2-10~keV}$ correlation. 
After completion of the {\it Chandra} AO8 observing cycle, X-ray information will be available 
for the most extreme radio-quiet Type~2 quasars from the Zakamska et al. (2003) sample.}
\label{lxoiiiz}
\end{figure}

\item 
In Vignali, Alexander \& Comastri (2006, hereafter V06), 
the most up-to-date results on the SDSS Type~2 quasar population were presented. 
Using a combination of {\it Chandra} and XMM-{\it Newton} pointed and serendipitous 
observations (for a total of 16 sources; filled triangles in Fig.~\ref{lxoiiiz}), 
selected predominantly among the most luminous in \hbox{[O\ {\sc iii}}], V06 detected X-ray 
emission from ten sources. 
For seven of these AGN, basic/moderate-quality X-ray spectral analyses constrained the 
column densities in the range \hbox{$\approx$~10$^{22}$ -- a~few~10$^{23}$~cm$^{-2}$} 
(filled triangles in Fig.~\ref{lx_compa}). Once their observed X-ray luminosities are corrected for 
the effect of absorption, there is indication that the X-ray luminosity predictions based on 
the Mulchaey et al. (1994) correlation are consistent with the values obtained from X-ray spectral 
fitting (all these seven sources lie close to the 1:1 line in Fig.~\ref{lx_compa}). 

\item 
%
Having calibrated the [O\ {\sc iii}] line luminosity as an indicator of the intrinsic X-ray emission 
on the seven sources with good X-ray photon statistics, 
there are indications that the X-ray undetected sources and the sources with a limited number of 
counts (open circles in Fig.~\ref{lx_compa}) 
are possibly more obscured than those found absorbed through direct X-ray spectral fitting, 
as pointed out also by Ptak et al. (2006). 
This would imply that up to $\approx$~50\% of the population 
is characterized by column densities in excess to $10^{23}$~cm$^{-2}$, with a sizable number 
of Compton-thick quasars possibly hiding among the X-ray faintest sources 
(see the quasars located in the grey region in Fig.~\ref{lx_compa}). 
This possibility is also suggested by the comparison of the X-ray-to-\hbox{[O\ {\sc iii}}]  
flux ratios of our sources vs. those obtained from a large sample of mostly nearby 
Seyfert~2 galaxies (see Guainazzi et al. 2005 and Fig.~6 of V06). 

\begin{figure}
\centering
\includegraphics[angle=0,width=0.55\textwidth]{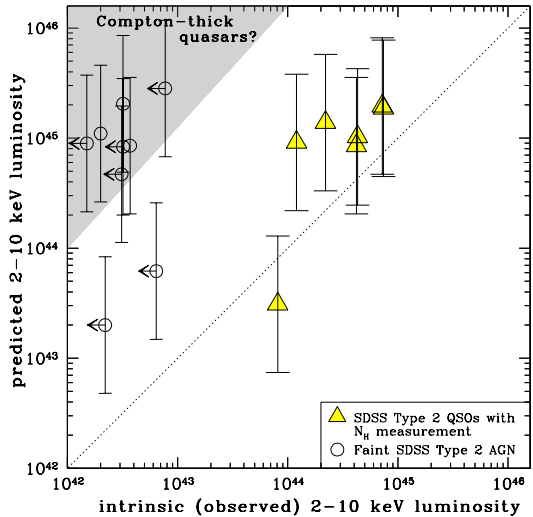}
\caption{
Comparison of the 2--10~keV luminosity computed from the compilation of V06 with that predicted 
assuming the Mulchaey et al. (1994) correlation. The dotted line shows the 1:1 ratio between the 
two luminosities. For the seven sources for which X-ray spectral fitting was possible using 
either {\it Chandra} or XMM-{\it Newton} data, 
the X-ray luminosity has been de-absorbed assuming the best-fit 
column density (filled triangles), while for the remaining X-ray fainter AGN (open circles), some of 
which undetected, the X-ray luminosity is derived from the X-ray flux with no correction for 
the unknown absorption. 
Leftward arrows indicate upper limits on the observed X-ray luminosity; the grey region shows the 
locus of heavily obscured, candidate Compton-thick quasars, where the observed luminosity is less 
than 1\% of the predicted one.}
\label{lx_compa}
\end{figure}

\item 
Given the highly inhomogeneous selection and incompleteness of the sample presented by Zakamska et 
al. (2003), the number density of SDSS selected Type~2 quasars can be derived only roughly. 
In an attempt to provide a first-order estimate, V06 obtained a value of $\approx$~0.05~deg$^{-2}$, 
%
%
while Gilli et al. (2007) XRB synthesis models predict a surface density of Type~2 quasars of 
$\approx$~0.15~deg$^{-2}$ in the $\approx$~0.3--0.8 redshift range. 
%
%
This comparison indicates that the Zakamska et al. (2003) selection is relatively efficient at 
finding obscured quasar activity; however, a combination of blank-field \hbox{X-ray} surveys and 
optical selection techniques will provide a more complete census. 
\end{list}

Given the accurate analyses carried out by Zakamska et al. (2003) in the original selection of 
the obscured SDSS AGN sample, it seems unlikely that a significant population of starburst 
galaxies or low-luminosity AGN is hiding among the sources with the highest \hbox{[O\ {\sc iii}}] 
luminosity in the V06 sample. Hence, it is likely that some of the Zakamska et al. (2003) 
quasars are obscured by Compton-thick material in the X-ray band. 
Although effective and comparatively complete, the optical selection and requirement for [OIII] 
in the SDSS spectra clearly limit obscured quasar searches to $z<1$. Due to the likely optical 
faintness of obscured quasars at $z>1$, a multi-wavelength selection process including targeted 
follow-up observations is likely to be required for a comparably complete census of 
obscured quasar activity at high redshift.

\section{What's next in the study of SDSS Type~2 quasars?}
The observations approved for {\it Chandra} Cycle~8 (10~ks pointing for 12 targets) will allow us 
to have an almost complete coverage of the most extreme SDSS Type~2 quasar candidates 
(i.e., above a predicted 2--10~keV 
luminosity of 10$^{44}$~erg~s$^{-1}$ also taking into account the dispersion in the 
$L_{[O\ III]}$--$L_{\rm 2-10~keV}$ correlation; open squares in Fig.~\ref{lxoiiiz}). 
While direct X-ray spectral fitting will be possible only for the most X-ray luminous SDSS 
Type~2 quasars, for the newly observed faint X-ray sources, basic spectral analysis will be 
carried out by means of the hardness-ratio technique. 
For the few-count or undetected sources, it will be possible to derive the average source 
properties through stacking analysis; 
use of this has been prevented thus far by the limited number of sources with 
either {\it Chandra} or XMM-{\it Newton} constraints and the paucity of X-ray counts. 
Luckily, it will be possible also to estimate the column density distribution of SDSS Type~2 quasars 
and, by stacking the X-ray source spectra in different $N_{\rm H}$ bins, to search for faint 
spectral features, as shown by Alexander et al. (2005b) for a sample of submillimeter galaxies.

\acknowledgements 
The authors acknowledge support by the Italian Space Agency (contract ASI--INAF I/023/05/0; 
CV and AC) and the Royal Society (DMA). 
The authors wish to thank the people involved in the HELLAS2XMM survey for useful discussions.


\end{document}